\documentstyle[preprint,aps]{revtex}
\begin{document}
\draft
%
%
\title{Magnetic Raman Scattering in Two-Dimensional\\
Spin-1/2 Heisenberg Antiferromagnets:
Spectral Shape Anomaly and Magnetostrictive Effects }
\author{Franco Nori,$^1$ R.~Merlin,$^1$ Stephan Haas,$^2$
Anders W.~Sandvik,$^2$ and Elbio Dagotto$^2$
}
\address
{
$^1$Department of Physics, The University of Michigan, Ann Arbor, MI 48109-1120
\\
$^2$Department of Physics,
Supercomputer Computations Research Institute
and National High Magnetic Field Lab,
Florida State University, Tallahassee, FL 32306
}
\date{\today}
\maketitle
\begin{abstract}
We calculate the Raman spectrum of the two-dimensional (2D) spin-1/2 Heisenberg
antiferromagnet by exact diagonalization and quantum Monte Carlo techniques on
clusters of up to 144 sites and, on a 16-site cluster, by considering the
phonon-magnon interaction which
leads to random fluctuations of the exchange integral.
Results are in good agreement with 
experiments on various high-$T_c$ precursors,
such as La$_2$CuO$_4$ and YBa$_2$Cu$_3$O$_{6.2}$.
In particular, our calculations
reproduce the broad lineshape of the two-magnon peak, the asymmetry about
its maximum, the existence of spectral weight at high energies, and the
observation of nominally forbidden $A_{1g}$ 
scattering.
\end{abstract}
\narrowtext

{\it Introduction.---\/}
Raman scattering is a powerful technique to study electronic excitations
in strongly correlated systems.  Recently, much attention has been
given to the anomalous 
magnetic scattering with a very broad and asymmetric lineshape
observed in the Raman spectra of the parent insulating compounds
of high-$T_c$ superconductors, such as La$_2$CuO$_4$,
and YBa$_2$Cu$_3$O$_{6.2}$ at around 3230cm$^{-1}$ and
3080cm$^{-1}$, respectively \cite{experiment}.
The selection rules associated with this peak are also anomalous.
While the spin-pair excitations scatter predominantly in the allowed
$B_{1g }$ channel, there is also a significant contribution in the nominally
forbidden $A_{1g}$ 
configuration, as well as much weaker $B_{2g}$ and $A_{2g}$
scattering\cite{experiment}.


Previous theoretical studies
on the spin-1/2 Heisenberg model for 2D square lattices
have computed the Raman spectra and its moments
for a nearest-neighbor interaction\cite{elbio,theory,liu,canali}
and only the moments 
when spin interactions along the plaquette diagonal were also
included.\cite{nnn}
These show good agreement with experiments in regard to the position of
the two-magnon peak, but they fail to account for the
spectral {\it shape}, and its enhanced {\it width}.

Several schemes have been considered to resolve this problem.  Initially,
from the analysis of the moments
it was proposed that strong quantum fluctuations were responsible for the
broadening (see, e.g., Ref.~\cite{elbio,nnn}).
However, recent studies of
spin-pair excitations in a {\it spin}--1 insulator, NiPS$_3$,
show a width comparable to that of the spin--1/2 
cuprates \cite{spin1}.  This questions the view
that the observed anomaly is due to large quantum fluctuations intrinsic
to spin--1/2 systems.
\ We remark that the measured widths are 3-4 times larger\cite{weber}
than those predicted by Canali and Girvin \cite{canali} within
spin-wave theory using the Dyson-Maleev transformation,
even when processes involving up to four magnons are taken into
account. The work by Canali and Girvin \cite{canali} and other
groups\cite{4-spin,marville} present convincing evidence that the
observed anomalous features of the magnetic scattering {\it cannot\/} be
satisfactorily explained by only considering quantum fluctuations.


In order to explain the observed anomalously broad and asymmetric
lineshapes, it seems then necessary to invoke an additional process.
Here, we consider the interaction between 
magnon pairs and phonons \cite{magnon-phonon}.
This mechanism is motivated in part by recent experimental observations
of a strong broadening of the $B_{1g}$ and an enhancement of the $A_{1g}$
scattering with increasing temperature\cite{knoll}.
In our approach we consider
the phonons as static lattice distortions which induce changes,
$\delta J_{ij} $,
in the exchange integral $J$ of the undistorted lattice.
We calculate the Raman spectra for a {\it nearest-neighbor\/}
Heisenberg model using a {\it nearest-neighbor\/} Raman operator
in the quenched-phonon approximation which, like the Born-Oppenheimer approach,
focuses on the fast (high-energy) magnon modes and freezes the slow
(low-energy) phonons.
This approximation is valid for the cuprates because there is a clear
separation of energies between the magnetic and vibrational modes.
For instance, in YBa$_2$Cu$_3$O$_6$
the characteristic Debye frequency is about 340cm$^{-1}$ while the two-magnon
excitation is $\approx$ 3080cm$^{-1}$.


{\it Raman Lineshape without phonon-magnon coupling.---}
The isotropic Heisenberg Hamiltonian is given by
$
\ H_0 = J \sum_{<ij>}  {\bf S}_i \cdot {\bf S}_j  \ ,
$
where the notation is standard, and only nearest neighbor interaction
is assumed. For the cuprates, the exchange
integral is $J \simeq 1450K \simeq 0.12$eV.
\ In our study, we obtained the ground state $| \phi_0 \rangle$
of $H_0$ on finite 2D square clusters with $N$ spins and periodic
boundary conditions using a Lanczos ($N=16, 26$), and Quantum Monte Carlo
(QMC) ($N=144$) algorithms.
We studied zero and finite temperature spectra associated with the
{\em nearest}-neighbor scattering operator [1-4]
\begin{equation}
R = \sum_{<ij>}
({\bf E}_{inc} \cdot {\bf \widehat{\sigma}}_{ij} )
({\bf E}_{sc} \cdot {\bf \widehat{\sigma}}_{ij} )
{\bf S}_i \cdot {\bf S}_j ,
\end{equation}
where ${\bf E}_{inc,sc}$ corresponds to the electric field
of the incident and scattered photons, and
${\bf \widehat{\sigma}}_{ij}$ is the unit vector connecting sites $i$
and $j$.
In the cuprates, and for nearest-neighbors only, the irreducible
representations of $R$ are $B_{1g}$, $A_{1g}$, and $E$.
We concentrate mainly on the dominant $B_{1g}$ scattering, e.g.,
 ${\bf E}_{inc} \propto \widehat{x} + \widehat{y}$ and
 ${\bf E}_{sc}  \propto \widehat{x} - \widehat{y}$.

The spectrum of the scattering operator can be written as
\begin{equation}
I(\omega) = \sum_n | \langle \phi_n | R | \phi_0 \rangle |^2
\delta (\omega - (E_n - E_0)) ,
\end{equation}
where $\phi_n$ denotes the eigenvectors of the Heisenberg model
with energy $E_n$. When doing exact diagonalizations on small clusters,
the dynamical spectrum $I(\omega)$ is extracted from a continued
fraction expansion of the quantity
\begin{equation}
I(\omega) = - \frac{1}{\pi} Im \langle \phi_0 |
R \frac{1}{\omega + E_0 + i\epsilon - H_0} R  | \phi_0 \rangle ,
\end{equation}
where $\epsilon$ is a small real number introduced in the
calculation to shift the poles of Eq.~3 into the complex plane.
\ In the QMC simulations, the imaginary-time correlator
$ \langle R(\tau) R(0) \rangle $ is calculated and $I(\omega)$
is obtained by numerically continuing this function to real
frequencies using a maximum entropy procedure\cite{gubernatis}.

Our calculated $B_{1g}$ spectra are shown in Fig.~1(a).
They were obtained from exact diagonalization ($N=16$) and QMC ($N=144$)
studies of the Heisenberg Hamiltonian on square lattices.
The two-magnon excitation observed experimentally lies around $3J$,
which is in good agreement with the location of the main peak
obtained from exact diagonalization in Fig.~1.
The position of this peak can be understood in terms of the Ising
model, which corresponds to the limit of the anisotropic Heisenberg
Hamiltonian when no quantum fluctuations are present.  In its ground state,
the Ising spins align antiferromagnetically for $J > 0$.  Within this model
and for a 2D square lattice, the incoming light creates a local spin-pair flip
at an energy $3J$ 
higher than the ground state energy.
This argument remains approximately valid even in the presence of quantum
fluctuations\cite{elbio,theory,liu,canali}.
Our results indicate that the two-magnon excitation is at
$2.9757J$,
$3.0370J$,
and $~3.2J$ for the
16-, 26-, and 144-site square lattices, respectively.
Finite-size effects are small because of the local nature of the Raman
operator.
For the 144-site lattice, the QMC calculation was carried out at a
temperature $T=J/4$.  The slight shift of the peak position,
compared to the $T=0$ results for the smaller clusters, is
consistent with the finite-$T$ exact diagonalization results of
Ref.\cite{theory}.  Statistical errors, absent in the exact diagonalization
results but unavoidable in any stochastic simulation, enhance the width of the
144-site spectrum.
These results confirm that neither finite-size effects nor finite temperature
can account for the discrepancies with the experimental spectra.

{\it Lineshape Anomaly.---\/}
The Raman spectra obtained from the pure Heisenberg model (see Fig.~1)
shows good agreement with experiments in regard to the two-magnon peak
position, but the calculated width is too small.
We will consider here the coupling between the magnon pair and phonons
\cite{magnon-phonon,knoll} to account for the observed wide and asymmetric
lineshape.
%
Our mechanism relates to that proposed by Halley \cite{halley} to account
for two-magnon infrared absorption in, e.g., MnF$_2$.

Quantum and thermal fluctuations distort the lattice.
\ The exchange coupling, which depends on the instantaneous positions of
the ions, can be expanded in terms of the their displacements
from equilibrium ${\bf u}$.
\ Keeping only the dominant linear terms:
$ J_{ij}({\bf r}) = J_{ij} =
J + \delta J_{ij} = J + {\bf u} \cdot \nabla J_{ij} ({\bf R}) $.
Here,
$\delta J_{ij} $ represents the instantaneous value of
${\bf u} \cdot \nabla J_{ij} ({\bf R})$, where
${\bf R}$ denotes the equilibrium position of the ion carrying the spin
(located at ${\bf r=R+u}$).
\ In the quenched-disorder approximation, the effective Hamiltonian is
\begin{equation}
H_1 =  \sum_{<ij>} (J + \delta J_{ij}) {\bf S}_i \cdot {\bf S}_j ,
\end{equation}
where $| \delta J_{ij} | < J$ is a random variable
corresponding to taking a snapshot of the lattice.
This new Hamiltonian is no longer translational invariant.


In our study, the random couplings $\delta J_{ij}$ were drawn from a Gaussian
distribution
$P(\delta J_{ij}) =
\exp{(- (\delta J_{ij})^2 / 2 \sigma )} / \sqrt{2 \pi \sigma } $.
$I(\omega )$ was obtained as the quenched average over
$m \simeq 1000$ realizations of the randomly distorted lattice.
The quenched average of an operator $\hat{O}$ is defined by
$ \langle \langle \hat{O} \rangle \rangle =
\frac{1}{m} \sum_{j=1}^m \langle \phi_0 (j)
|\hat{O} |\phi_0 (j) \rangle ,
$
where $\phi_0 (j)$ is the ground state of the $j$th realization of
the disordered system.


In Fig.~1(b) we show the $B_{1g}$ Raman spectrum from Eq.~(1)
for a 16-sites square lattice 
with $\sigma \sim 0.4J$, which we found to agree best
with experimental spectra \cite{experiment}.
Our calculations do not consider the effect of frozen phonons on the scattering
operator $R$.  Notice that the coefficients pertaining to $R$ are generally
unrelated to the matrix elements of the system's Hamiltonian
(e.g., $\partial J/ \partial Q$ in $H$ bears on $e^2/r$,
while the corresponding terms $\propto Q S_i S_j$ in $R$ bear on the dipole
moment). 
In particular, and unlike the case without phonons, the fully symmetric
$A_{1g}$ component of the scattering operator does not commute with $H$.

We find that the three main features observed in the $B_{1g}$
configuration\cite{experiment}, namely, the broad lineshape of the
two-magnon peak, the asymmetry about its maximum, and the existence of
spectral weight up to $\omega \sim 7J$ are well reproduced.
Beyond the two-magnon peak, there is a continuum of
phonon-multi-magnon excitations. The small feature around
$\omega \simeq 5.5J$ (for $0 \leq \sigma \leq 0.3J$) is compatible with 
a four-magnon excitation.

%

{\it Magnetostriction.---\/}
Since the effects of the phonon-magnon interaction (i.e., magnetostriction)
have not been extensively studied by theoretical work in the cuprates,
a few comments are in order.
\ The coupling between the spin and strain degrees of freedom
modifies both elastic and magnetic properties. 
In fact, there are extensive studies on the (sometimes very strong) influence
of elasticity on
magnetism\cite{mattis-shultz,striction-1,recent-prls}.
Mattis and Shultz\cite{mattis-shultz} considered the influence of
{\it uniform compression\/} (i.e., all bonds {\it equally distorted})
in their classic study of magnetothermomechanics.  Their results were
criticized\cite{striction-1} for ignoring
the effects of phonons (i.e., {\it local fluctuations\/} in the bond lengths,
which are taken into account in the present work).
Recently, {\it giant\/} magnetostrictive effects have been
reported in several high-$T_c$ superconductors\cite{striction-htcsc}.
Also, important magnetostrictive effects have been reported in
heavy-fermion\cite{striction-hf} and
low-$T_c$\cite{striction-ltc} superconductors.

{\it Superexchange-Phonon Coupling.---\/}
The width of the Gaussian distribution, $\sigma$, represents changes in $J$
due to large $incoherent$ atomic displacements.  Thus, one can write
$\sigma \sim |\langle \delta \ln J / \delta Q \rangle \langle Q \rangle |$
where $\langle Q \rangle$ is an average {\em zero point motion} (at $T=0$)
and $ \langle \delta J / \delta Q \rangle $
is a weighted average of $\nabla J_{ij}$ with respect to the displacement
of {\it all\/} the ions participating in the exchange.
Parenthetically, it is trivial to treat the case $T \neq 0$ by increasing
$\delta J$.
Let $r$ be the Cu-Cu distance, $\upsilon$ the sound velocity,
and $M$ an effective reduced mass for the ions.
A simple calculation gives
$ \langle Q \rangle / r \sim (M \upsilon r / \hbar)^{-1/2} \sim 0.05 $
which is consistent with X-ray measurements of the mean displacement
of oxigen atoms normal to the layers\cite{d1,d2}.
While $\nabla J_{ij}$ is not known for most phonons, values for longitudinal
acoustic modes can be gained from the $r$-dependence of $J$ in the form
$J(r) \sim r^{-\alpha}$ or
$\partial \ln J / \partial \ln r = - \alpha $\cite{Jpressure}.
For conventional transition metal oxides and halides,
$10 \leq \alpha \leq 14$\cite{Jpressure},
in reasonable agreement with the theoretical estimate
$\alpha=14$\cite{harrison}.
\ For the cuprates, high-pressure Raman measurements\cite{aronson}
and material trends\cite{lance} give, respectively,
$\alpha \approx 5-7$ and $\alpha \approx 2-6$.  These values translate
into $\sigma \approx (0.1-0.35)J$.
We emphasize that the relevant {\em incoherent\/}
$\delta J$'s (or $\delta Q$'s) of our case are much larger
than those in pressure studies involving {\em coherent\/} motion of ions
(see, e.g., the discussion in p.~466 of \cite{halley}).  Thus, we must use
larger $\sigma$ ($\sigma \sim 0.4J$).

Finally, we would like to stress that not every kind of disorder
gives rise to the observed broadening of the spectrum.
For instance, disorder by point defects or twinning planes will not
produce such an effect.  Also, it is observed in experiments that the Raman
linewidth broadens with increasing temperature\cite{knoll}. This is a strong
indication of a phonon mechanism for the broadening.


{\it $A_{1g}$ and $B_{2g}$ Symmetries.---\/}
For the $A_{1g}$ symmetry, the undistorted Raman operator commutes with the
Heisenberg Hamiltonian, and {\it no\/} scattering can take place.  However,
the addition of disorder changes the commutator and can produce an $A_{1g}$
signal.  Instead, the silent $B_{2g}$ channel remains forbidden within our
{\em nearest}-neighbor Raman operator.
\ Fig.~1(c) shows the comparison between our numerically obtained $A_{1g}$
spectra (for $\sigma \sim 0.4J$) and the experimental
results \cite{experiment,nnn}.
\ The agreement between theory and experiments is reasonably good.
We stress that the $A_{1g}$ scattering follows naturally from our model
unlike approaches relying on additional hypotheses, like, for instance,
diagonal-nearest-neighbor couplings\cite{nnn},
4-spin terms\cite{4-spin},
new fermionic quasiparticles\cite{hsu}, or spinons.
\ For a detailed discussion of these and other proposed explanations
of the lineshape anomaly, see \cite{canali,marville}.

{\it Extensions.---\/}
The mid-infrared optical absorption in undoped lamellar copper oxides
show broad features which   
are believed to originate from exciton-magnon absorption
processes\cite{graybeal}.
Instead, these results could be interpreted as due to the interaction between
phonons and magnons\cite{infrared}.
Also, excellent agreement with conductivity experiments has been recently
achieved by the inclusion of phonon-induced strong-disorder\cite{fehrenbacher}.


{\it Summary.---\/} We find that light scattering spectra by spin
excitations is caused by intrinsic spin-spin interactions and by
interactions with phonons.  We provide strong evidence that
the two-magnon Raman peak is strongly modified by coupling to
low-energy phonons which randomly distort the lattice.
Our calculations are in good agreement with experiments and
provide a simple explanation of four puzzling features of the data:
the broad lineshape of the two magnon peak, the asymmetry about its maximum,
the existence of a spectral weight at high energies, and the observation of
nominally forbidden $A_{1g}$ scattering.

We thank J.~Riera for his help. 
S.H. acknowledges support by
the Supercomputer Computations Research Institute (SCRI).
The work of E.D. and R.M. is partially supported by
the donors of The Petroleum Research Fund, and the work of E.D. and A.S.
by the ONR grant N00014-93-1-0495.





\begin{figure}
\caption{
Normalized Raman cross section, $I(\omega)/I_{max}$,
versus $\omega /J$,
for the spin-1/2 Heisenberg model with $N$ sites.
(a) $B_{1g}$ Raman spectra  
obtained from exact diagonalization with $N=16$, $T=0$, and $\epsilon=0.1J$;
and from QMC with $N=144$ and $T=J/4$.
$B_{1g}$ (b) and $A_{1g}$ (c) spectra obtained from exact diagonalization
($N=16$) with randomness in the exchange
integral representing the interaction between spin-pairs and the phonons.
The continuous, dashed and dotted lines in (b) and (c) correspond,
respectively, to $\sigma=0.3J$, $0.4J$, and $0.5J$.
\ For comparison, the experimental results (from Ref.[6]) are shown.
}
\label{fig1}
\end{figure}


\begin{references}

\bibitem{experiment}
 K.B.~Lyons {\it et al.},
Phys.~Rev.~B {\bf 39}, 2293 (1989);
I.~Ohana      {\it et al., ibid} {\bf 39}, 2293  (1989);
P.E.~Sulewski {\it et al., ibid} {\bf 41}, 225   (1990);  
T.~Tokura     {\it et al., ibid} {\bf 41}, 11657 (1990);  
S.~Sugai      {\it et al., ibid} {\bf 42}, 1045  (1990).



\bibitem{elbio}E.~Dagotto and D.~Poilblanc, Phys.~Rev.~B {\bf 42}, 7940 (1990).

\bibitem{theory}
F.~Nori, E.~Gagliano, and S.~Bacci, Phys.~Rev.~Lett.~{\bf 68}, 240 (1992);
S.~Bacci and E.~Gagliano, Phys.~Rev.~B {\bf 43}, 6224 (1991);
{\bf 42}, 8773 (1990).

\bibitem{liu} Z.~Liu and E.~Manousakis, Phys. Rev. B {\bf 43}, 13246 (1991).

\bibitem{canali}C.M. Canali and S.M. Girvin,
Phys.~Rev.~B {\bf 45}, 7127 (1992).

\bibitem{nnn}
R.R.P.~Singh {\it et al.}, Phys.~Rev.~Lett.~{\bf 62}, 2736 (1989).


\bibitem{spin1}
S.~Rosenblum, A.H.~Francis, and R.~Merlin, Phys.~Rev.~B {\bf 49}, 4352 (1994).

\bibitem{weber} H.W.~Weber and G.W.~Ford, Phys.~Rev.~B {\bf 40}, 6890 (1989).

\bibitem{4-spin}
S. Sugai,
Sol.~Stat.~Comm.~{\bf 75}, 795 (1990);
%
M.~Roger and J.M.~Delrieu,
Synthetic Metals {\bf 29}, F673 (1989).

\bibitem{marville} L.~Marville, 1992 Ph.D.~MIT thesis.

\bibitem{magnon-phonon}
M.J.~Massey,R.~Merlin, and S.M.~Girvin, Phys. Rev. Lett.
{\bf 69}, 2299 (1992);
J.B.~Sokoloff, J.~Phys.~C {\bf 5}, 2482 (1972);
M.G.~Cottam, J.~Phys.~C {\bf 7}, 2901 (1974).

\bibitem{knoll} P.~Knoll {\it et al.}, Phys.~Rev.~B {\bf 42}, 4842 (1990).

\bibitem{gubernatis} J.E.~Gubernatis {\em et al.},
Phys.~Rev.~B {\bf 44}, 6011 (1991).

\bibitem{halley} J.W.~Halley, Phys.~Rev.~{\bf 154}, 458 (1967).

\bibitem{mattis-shultz}
D.C.~Mattis and T.D.~Shultz, Phys. Rev. {\bf 129}, 175 (1963).

\bibitem{striction-1}
M.E.~Fisher, Phys.~Rev.~{\bf 176}, 257 (1968);
G.A.~Baker and J.W.~Essam, Phys.~Rev.~Lett.~{\bf 24}, 447 (1970);
H.~Wagner and J.~Swift, Z.~Phys.~{\bf 239}, 182 (1970);
H.C.~Bolton and B.S.~Lee, J.~Phys.~C {\bf 3}, 1433 (1970);
F.J.~Wegner, {\it ibid} 
{\bf 7}, 2109 (1974);
%
E. Pytte, Phys.~Rev.~B {\bf 10}, 2039 (1974).
%
D.J.~Bergman and B.I.~Halperin, Phys. Rev. B {\bf 13}, 2145 (1976);
Z.-Y.~Chen and M.~Kardar, {\it ibid} 
{\bf 30}, 4113 (1984).

\bibitem{recent-prls} M.~Hase, {\it et al.,} 
Phys.~Rev.~Lett.~{\bf 70}, 3651 (1993);
J.P.~Pouget, {\it et al.,} 
{\it ibid} {\bf 72}, 4037 (1994);
K.~Hirota, {\it et al.}, 
{\it ibid} {\bf 73}, 736 (1994);
Q.J.~Harris, {\it et al.}, 
Phys.~Rev.~B {\bf 50}, 12606 (1994).

\bibitem{striction-htcsc} 
H.~Ikuta, et al. 
Phys.~Rev.~Lett.~{\bf 70}, 2166 (1993);
A. del Moral, et al., 
Physica C {\bf 161}, 48 (1989);
A.M.~Kadomtseva, et al.,
{\it ibid} {\bf 162}, 1361 (1989).

\bibitem{striction-hf} 
F.E.~Kayzel, et al., 
Physica B {\bf 147}, 231 (1987);
A. de Visser, et al., 
{\it ibid}, {\bf 165-166}, 375 (1990).

\bibitem{striction-ltc} 
M.~Isino et al., 
Phys.~Rev.~B {\bf 38}, 4457 (1988).


\bibitem{d1} G.H.~Kwei, A.C.~Lawson, M.~Mostoller,
Physica C {\bf 175}, 135 (1991).

\bibitem{d2} A.S. Borovik, 
A.A.~Epiphanov, V.S.~Malyshevsky, and V.I.~Makarov,
Phys.~Lett.~A {\bf 161}, 523 (1992).

\bibitem{Jpressure}
see, e.g., M.J.~Massey {\it et al.}, Phys.~Rev.~B {\bf 41}, 8776 (1990),
and references therein.
%

\bibitem{harrison}
W.A.~Harrison, {\it Electronic Structure and the Properties
of Solids}  (Freeman, San Francisco, 1980).

\bibitem{aronson}
M.~Aronson {\it et al.}, 
Phys.~Rev.~B {\bf 44}, 4657 (1991).

%
%

\bibitem{lance}
S.L.~Cooper {\it et al.,} 
Phys.~Rev.~B {\bf 42}, 10785 (1990).



\bibitem{hsu} T.~Hsu, Phys.~Rev.~B {\bf 41}, 11379 (1990);
this Raman lineshape does not agree with experiments, e.g.,
the $A_{1g}$ and $B_{1g}$ spectra look almost identical, and
the asymmetry has the wrong sign.

\bibitem{graybeal} J.D.~Perkins et al
Phys.~Rev.~Lett. {\bf 71}, 1621 (1993).

\bibitem{infrared} Merlin {\it et al.} (unpublished);
J.~Lorenzana and G.A. Sawatzky, (unpublished);
J.R.~McBride, L.R.~Miller, and W.H.~Weber,
Phys.~Rev.~B {\bf 49}, 12224 (1994);
B.~Normand, H.~Kohno, and H.~Fukuyama (unpublished).

\bibitem{fehrenbacher} R.~Fehrenbacher, Phys.~Rev.~B {\bf 49}, 12230 (1994).


\end{references}
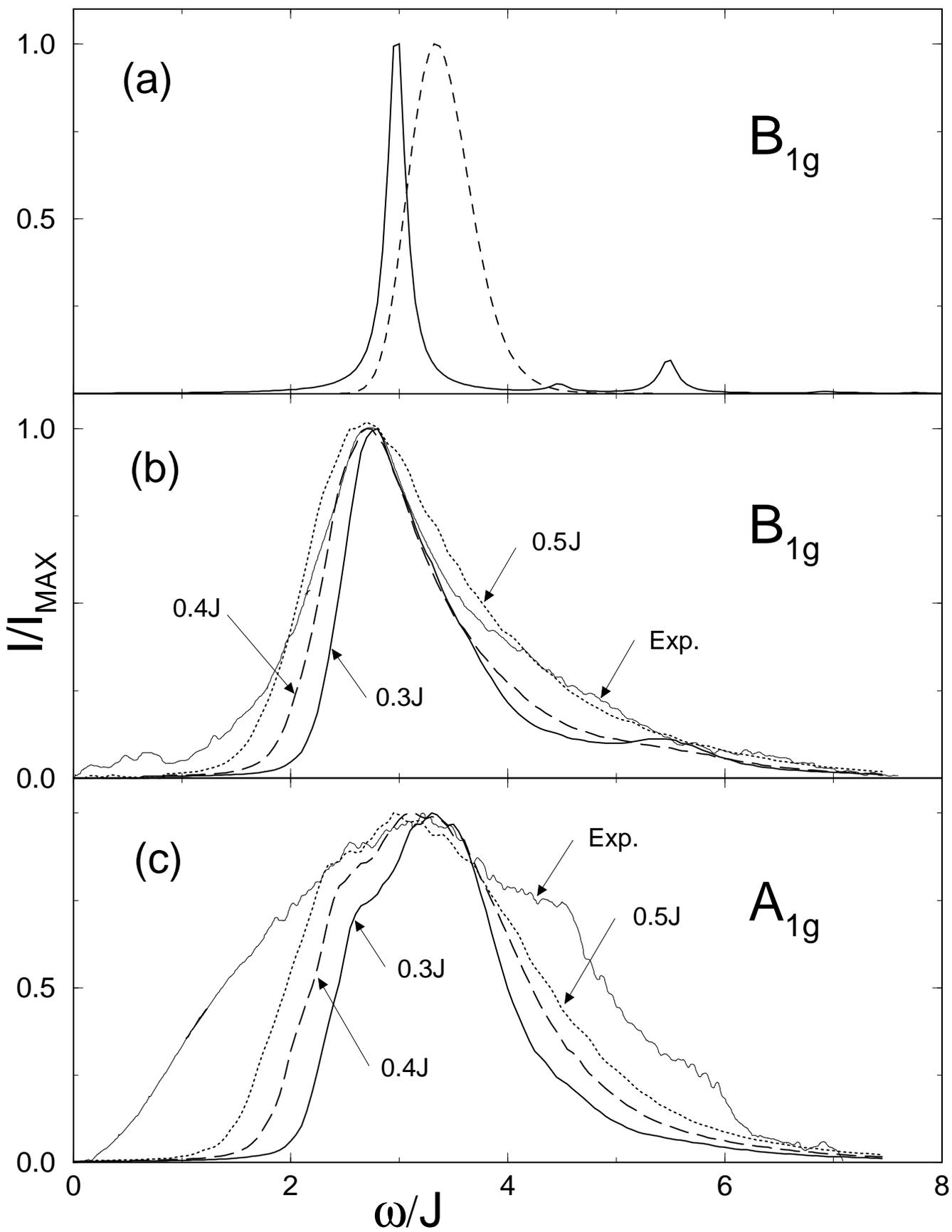
\end{document}